# Symmetry broken Chern insulators and "magic series" of Rashba-like Landau level crossings in magic angle bilayer graphene

Ipsita Das[1†], Xiaobo Lu[1†*], Jonah Herzog-Arbeitman[2], Zhi-Da Song[2], Kenji Watanabe[3], Takashi Taniguchi[3], B. Andrei Bernevig[2] and Dmitri K. Efetov[1*]

1. ICFO - Institut de Ciencies Fotoniques, The Barcelona Institute of Science and Technology, Castelldefels, Barcelona 08860, Spain.
2. Department of Physics, Princeton University, Princeton, New Jersey 08544, USA
3. National Institute for Materials Science, 1-1 Namiki, Tsukuba 305-0044, Japan

*Correspondence to: xiaobo.lu@icfo.eu and dmitri.efetov@icfo.eu; †These authors contributed equally to this work.

**Flat-bands in magic angle twisted bilayer graphene (MATBG) have recently emerged as a rich platform to explore strong correlations, superconductivity and magnetism. However, the phases of MATBG in magnetic field, and what they reveal about the zero-field phase diagram remain relatively unchartered. Here we use magneto-transport and Hall measurements to reveal a rich sequence of wedge-like regions of quantized Hall conductance with Chern numbers $C = \pm1, \pm2, \pm3, \pm4$ which nucleate from integer fillings of the moiré unit cell $\nu = \pm3, \pm2, \pm1, 0$ correspondingly. We interpret these phases as spin and valley polarized Chern insulators, equivalent to quantum Hall ferromagnets. The exact sequence and correspondence of Chern numbers and filling factors suggest that these states are driven directly by electronic interactions which specifically break time-reversal symmetry in the system. We further study quantum magneto-oscillation in the yet unexplored higher energy dispersive bands with a Rashba-like dispersion. Analysis of Landau level crossings enables a parameter-free comparison to a newly derived "magic series" of level crossings in magnetic field and provides constraints on the parameters $w_0$ and $w_1$ of the Bistritzer-MacDonald MATBG Hamiltonian. Overall, our data provides direct insights into the complex nature of symmetry breaking in MATBG and allows for quantitative tests of the proposed microscopic scenarios for its electronic phases.**

Electron interactions can break the underlying symmetries of condensed matter systems and drive the formation of a multitude of exotic quantum phases. In twisted bilayer graphene rotated by the magic angle of $\theta_m \approx 1.1°$ (MATBG), ultra-flat moiré bands (1-6) host strongly interacting electrons which give rise to a plethora of phases including correlated insulators (CI)(1, 7-10), superconductors (SC)(2, 7, 11-21) and orbital magnets (OM)(7, 22-24). The multitude of phases observed in MATBG suggests the presence of many interesting low-energy states in the many-body spectrum, some of which can be stabilized by external knobs such as pressure, strain and fields. However, to this date the exact nature of the interaction-driven symmetry broken ground states is still not well understood. Recently it was proposed that the two spin and valley degenerate flat-bands can be understood as a series of 8-fold (quasi-)degenerate topologically non-trivial bands with opposite Chern numbers $C = 1$ and $C = -1$ (25-34). However direct experimental evidence of these states remains scarce (12, 31-33). Moreover, the nature of the high-energy dispersive bands and the possible phases that can appear at higher filling factors has not been explored.

To reveal the (topologically) non-trivial properties of the low-energy states of MATBG, it is necessary to gap them and lift their degeneracy. The single-particle band-structure of

MATBG obeys the inversion and time-reversal symmetries $C_2T$, which protects Dirac points. Breaking this symmetry can create a gap and give rise to Chern insulators with quantized Hall conductance. This has been achieved in the past by breaking the $C_2$ lattice symmetry at the single-particle level by structural alignment with hexagonal boron nitride (hBN) substrates (22-24). Recent studies also observed Chern insulators in devices without hBN alignment and hence without explicit single-particle symmetry breaking (7, 12). For these samples, strong electronic interactions could directly break one or both of the $C_2$ and $T$ symmetries and valley polarize the system, creating isolated (quasi-)flat-bands in the K and K' valleys and giving rise to a sequence of quantum Hall ferromagnets.

Here we report on detailed magneto-transport studies of several twisted bilayer graphene devices close to the firsts magic angle $\theta = 1.03° - 1.15°$, which were specifically not aligned to the insulating hBN substrates. Fig. 1a displays the schematic of the typical devices, which consist of a quadruple van der Waals hetero-structure of graphite/hBN/MATBG/hBN. We perform four-terminal longitudinal resistance $R_{xx}$ and Hall resistance $R_{xy}$ measurements, where the carrier concentration $n$ in the MATBG is continuously controlled by a voltage $V_g$ on the graphite gate. We normalize $n$ by $n_s = 4n_0$, the density of the fully filled 4-fold spin and valley degenerate moiré bands, and define the filling factor of carriers per moiré unit cell $\nu = n/n_0$. Applying a perpendicular magnetic field $B_\perp$ at $T = 1.5K$ in Fig. 1b and c, reveals a set of broad wedge-shaped regions in the $n$-$B$ phase-space where longitudinal resistance $R_{xx} \sim 0\ \Omega$ and $R_{xy} \sim h/Ce^2$ (where e is the electron charge, h is Planck's constant). These quantized regions follow a linear slope of $dn/dB = Ce/h$ which can be traced to different integer fillings $\nu$ at $B = 0$. These singular states show a clear correspondence between Chern number and filling factor $(C, \nu)$. We find a robust sequence for the $(\pm 4, 0)$, $(\pm 3, \pm 1)$, $(\pm 2, \pm 2)$ and $(\pm 1, \pm 3)$ states in different devices (12), as is summarized in Fig. 1e, which shows the schematic of their cumulative phase diagrams.

At lower $T$, we also observe a clear set of Landau level (LL) fans, which follow the typically reported 4-fold degeneracy at the charge neutrality point (CNP) (SI). Strikingly, the energy gaps associated with these states are up to an order of magnitude smaller than the gaps of the $(C, \nu)$ states, which have values of ~1meV (see SI). The $(C, \nu)$ states are more visibly pronounced, and some already quantize below $B_\perp < 0.3T$ and $T < 10K$ (SI), in contrast to LL quantization, which in MATBG occurs typically above $B_\perp > 3T$ and below $T < 1K$. This allows us to disentangle these gaps by elevating $T$, and to observe their robust and unperturbed sequence. However, these states do not form in zero $B_\perp$-field; even though they require a $B_\perp$-field to nucleate, the $(\pm 4, 0)$, $(3, 1)$ and $(-2, -2)$ already form at a negligibly small $B_\perp > 0.1T$. The very small values of the field for which these states appear, especially the $(-2,-2)$ state (12), suggests they are very competitive to the true ground-state in zero field. In contrast the $(-3, -1)$ and $(\pm 1, \pm 3)$ states require higher fields of $B_\perp > 2T$. Fig. 1e summarizes the corresponding critical $B_\perp$-fields for all the states and devices.

We interpret these states as correlated Chern insulators (CCI) stabilized by interactions and small $B_\perp$-field. These states are possible because the underlying flat-bands of MATBG can be thought of as Chern bands with 8-fold valley, spin and sub-lattice degeneracy, which carry opposite Chern numbers $C = 1$ and $C = -1$ (25-34). Lifting the degeneracy of these bands by gaping out their Dirac points and polarizing them can create topologically non-trivial gaps, in direct analogy to quantum Hall antiferromagnets (36-38). In order to explain the data, the breaking of degeneracy has to happen in a $T$-breaking rather than $C_2$ breaking fashion. Overall this picture can well explain the observed pecking order of the Chern numbers between neighboring states, which indicates that starting from $\nu = 0$, the total Chern number of $C = 4$ is decreased by 1 when ν is increased to the next integer value on the electron side. Equivalently, on the hole side, starting from $C = -4$ at ν = 0, $C$ increases by 1 when ν is decreased to the next integer value. This can be understood as the repeated filling of bands with Chern number $C = -1$, for electrons and $C = 1$ for holes, in the valley and spin sub-bands (31-33) (Fig. 1d).

The exact sequence of Chern numbers at different fillings gives insight into the dominant symmetry breaking mechanisms. Unlike in previous studies, where $C_2$ symmetry was deliberately broken by structural alignment of MATBG with the hBN substrates, we avoid this in our devices, by keeping the angle between the crystallographic orientations of MATBG and hBN $θ > 10°$. Generally, broken $C_2$ symmetry (and keeping $T$ unbroken) leads to a sign reversal of the Chern numbers in the K and K' valleys, while in contrast, broken $T$ symmetry (and keeping $C_2$ unbroken) is expected to preserve their Chern numbers. This can have a clear imprint on the resulting many-body states and the predicted Chern states for each ν. While breaking either symmetry predicts the (±4, 0), (±2, ±2) and (±1, ±3) states, $C_2$ breaking predicts the existence of the (±1, ±1), instead of the (±3, ±1) which is predicted from $T$ symmetry breaking and observed in the experiment (31-33). We suggest therefore that interactions in $B_⊥$-field in MATBG specifically break time reversal symmetry.

A central question is why the Chern insulators require a non-zero $B_⊥$-field to nucleate. Since the (±4, 0), (3, 1) and (-2, -2) states occur already in the weak field limit as low as $B_⊥ \sim$ 0.1T, which corresponds to a negligibly small magnetic flux per moiré unit cell of only $Φ < 0.01 Φ_0$ (where $Φ_0$ is the flux quantum), we find unlikely a mechanism associated with the onset of Hofstadter sub-bands (33). One strong hypothesis could be that the symmetry breaking many-body state which gives rise to the Chern insulators may already exist at $B_⊥ = 0$ but the phase is obscured by disorder and phase separation into domains until $B_⊥$-field aligns and stabilizes it (39). Another contending, more probable scenario is that at $B_⊥ = 0$ many-body states with Chern numbers closely compete with one another, including topologically trivial $C = 0$ states, where the application of $B_⊥$ favors states with higher $C$ values (42-44).

The symmetry and topology of the underlying ground state has dramatic implications for the understanding of the possible mechanisms responsible for all other emergent phases like correlated insulators (1, 7-10), superconductors (2, 7, 11-21), and orbital magnets (7, 22-24), which we also observe in the reported devices (Fig. 1e and Fig. 2a-c). Their phase-space lies close to the Chern insulators, where most samples display robust SCs and topologically trivial insulators around ν = ±2. In particular, device A2 shows a record high critical SC temperature for MATBG devices of $T_c \sim 5$K (Fig. 2b), and strong resistance peaks at half-filling which interrupt the SC domes (Fig. 2a). However, in the previously reported device D1 (12), we did not observe a $C = 0$ insulator at ν = -2, where instead we have found a broad SC dome region and a $C = -2$ CI which is formed above $B_⊥ > 0.2$T. In addition, in device A3 at ν = 1 we observe an OM at $B_⊥ \sim 2$T very close to a forming CI (Fig. 2c), likely created through a similar microscopic mechanism. Overall these findings may suggest a close competition between topologically trivial and non-trivial insulators at $B_⊥ = 0$, which directly impacts the SC and OM phases.

We further tune the carrier density beyond full-filling of the flat-bands |ν| > 4, and populate the yet largely unstudied higher energy dispersive bands of MATBG. Fig. 3a shows the calculated energy dispersion of MATBG using the Bistritzer-MacDonald model (3) including corrugations (see SI). The lowest lying two 4-fold degenerate higher energy bands can be modeled with a Rashba-like Hamiltonian, i.e. a two-band model to order $O(k^2)$ that transitions from a Dirac point into quadratic free electron bands (Fig. 3b). The model can be justified by a

perturbative calculation, and the trigonal-distortion terms are also computed (see SI). The Rashba-like Hamiltonian does not originate from spin-orbit coupling, which is negligible in these samples, but from crystalline terms. In this regime we perform $B_\perp$-field dependent $R_{xx}$ vs. $v$ measurements, which reveal a rich 4-fold degenerate LL spectrum with a multitude of well pronounced LL crossings (40,41), Hofstadter sub-bands and a new set of LLs at $|v| = 8$ (Fig. 3c). These features are well developed in all devices, and show relatively electron-hole symmetric features in the valence and conduction bands.

We exactly solve the Rashba Hamiltonian in $B_\perp$-field and obtain LLs (see SI) which are in excellent quantitative agreement with the experimental findings. The calculation reveals a series of LLs, each carrying Chern number $C = 4$, that interpolate between the Dirac node and quadratic free electron regimes. Critically, all the LLs undergo a series of crossings, or gaps closings, as is calculated (see SI) and experimentally seen in $R_{xx}$ vs. $v$ vs. $B_\perp$ measurements close to the band-gaps in Fig 3c, and in Fig. 3f which converts the same data for clarity into $R_{xx}$ vs. $C$ vs. $1/B_\perp$. The LL crossing between the $n^{th}$ and $(n+1)^{th}$ LLs happen at a well-defined field $B_\perp = B^*$ (Fig. 3d), and can be understood as an interruption of the LL gap with $C = n$ in the LL fan diagram, however the gap reappears again at a higher field (Fig. 3e). Each LL gap is interrupted several times, where we count the different generations of crossings for each LL from high to low $B_\perp$. We extract the $B^*$ values for all the LL crossing from Fig. 3c and f, normalize these to $B^*_{|24|,3}$, the field at the 3$^{rd}$ crossing of the LL with $C = \pm 24$ (for electrons and holes respectively), and fit them with our calculations in Fig. 3g. This allows us to extract an estimate of the Rashba coupling parameter $\xi$ (see SI). Moreover, we find that the ratios between all $B^*$ values are independent of all the parameters of the low energy Hamiltonian, and so present a stringent, parameter-free test of the physics – we call them the "magic series". The two corrugation parameters $w_0$ and $w_1$ of the Bistritzer-MacDonald Hamiltonian are constrained by the measured Rashba coupling $\xi = 0.186/\Omega$ (in units where $\hbar = 1$ and $\Omega = 2\sqrt{3}(13.5nm)^2$ is the area of the moiré unit cell), presenting a direct estimation of the physical parameters of MATBG (see SI). Neglecting the particle-hole symmetry breaking, we find that the ranges $0.7 \leq w_0/(v_F k_D/\sqrt{3}) \leq 0.8$ and $0.95 \leq w_1/(v_F k_D/\sqrt{3})$ give good agreement with the measurements of the Rashba coupling, where $v_F$ is the Fermi velocity and $k_D$ is the moiré wavevector (28,43,44).

In summary, our data provides a new and detailed view of the high $B_\perp$-field phase diagram of MATBG and demonstrates its underlying topologically non-trivial properties, which are revealed by interaction-driven symmetry breaking. Our experimental and theoretical analysis imposes stringent boundary conditions for microscopic scenarios for its complex electronic phases. Detailed analysis of the states in the upper passive bands revealed a remarkable agreement between theory and experiment, enabling a novel method of extracting the band parameters of the Bistritzer-MacDonald Hamiltonian.

Acknowledgements:
We are grateful for fruitful discussions with Ali Yazdani, Eva Andrei, Dima Abanin and Andrea Young. Funding: D.K.E. acknowledges support from the Ministry of Economy and Competitiveness of Spain through the "Severo Ochoa" program for Centres of Excellence in R&D (SE5-0522), Fundació Privada Cellex, Fundació Privada Mir-Puig, the Generalitat de Catalunya through the CERCA program, funding from the European Research Council (ERC) under the European Union's Horizon 2020 research and innovation programme (grant agreement No. 852927)" and the La Caixa Foundation. B.A.B. was supported by the Department of Energy Grant No. de-sc0016239, the Schmidt Fund for Innovative Research, Simons Investigator Grant No. 404513, and the Packard Foundation. Further support was provided by the National Science Foundation EAGER Grant No. DMR 1643312, NSF-MRSEC DMR-1420541, BSF Israel US foundation No. 2018226, ONR No. N00014-20-1-2303, and Princeton Global Network Funds. I.D. acknowledges the support from INphINIT "La Caixa" (ID 100010434) fellowship program (LCF/BQ/DI19/11730030).

**Supplementary Information** is available for this paper.

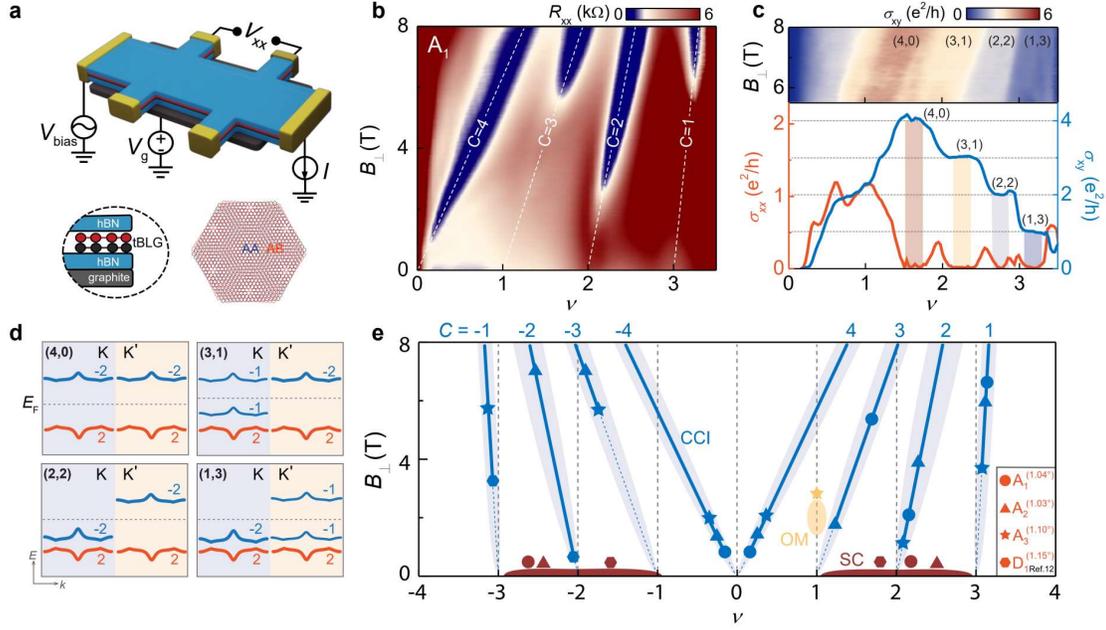

**Fig. 1. Emergent correlated Chern insulators in MATBG.** (**a**) Schematic of the hBN encapsulated and graphite-gated MATBG device. (**b**) Color plot of $R_{xx}$ vs. $\nu$ and $B_\perp$, measured at $T$ = 1.5K. White lines indicate the trajectories of four different topologically non-trivial Chern gaps with $(C, \nu)$ indices (4,0), (3,1), (2,2) and (1,3). (**c**) (upper panel) Corresponding Hall conductance $\sigma_{xy}$ vs. $\nu$ and $B_\perp$. (lower panel) Line-cuts showing quantized $\sigma_{xy}$ and vanishing longitudinal conductance of the Chern insulators at $B_\perp$ = 8T. (**d**) Schematic of the proposed band arrangement sequence for the symmetry broken topological bands and correlated Chern insulators at different fillings ($\nu$ = 0, 1, 2 ,3). We note that a valley coherent configuration for the (2,2) state is also allowed (Extended Data). (**e**) Cumulative phase diagram schematic of several devices, summarizing the observed correlated Chern insulators (CCI), superconductors (SC) and orbital magnets (OM) in the $\nu$-$B_\perp$ plane. The position of the device icons for each CCI corresponds to the critical $B_\perp$-field values at which these states form. Similarly, the optimal doping positions for SC are marked by corresponding device icons.

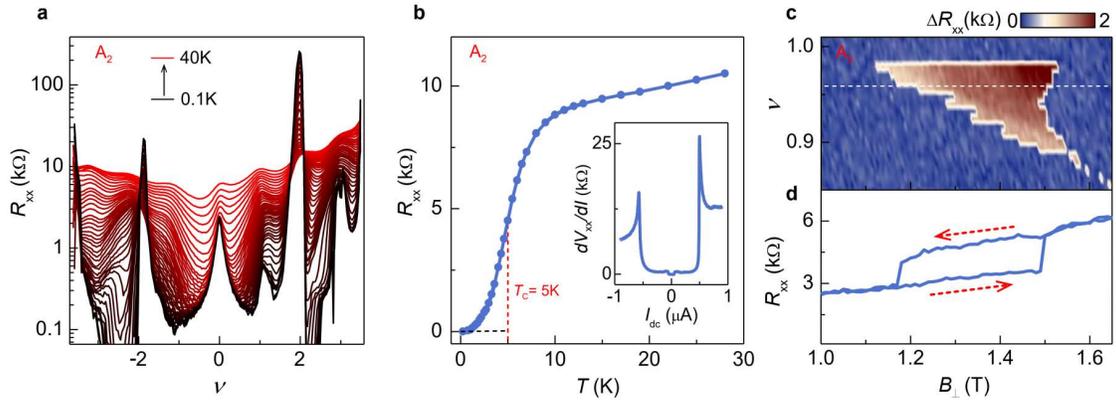

**Fig. 2. Correlated insulators, superconductors and orbital magnets in MATBG.** (**a**) $R_{xx}$ vs. $v$ of device A2 at different temperatures, shows well developed CI at $v = \pm 2$, as well as SC regions, which are shown in (**b**) Temperature dependence of $R_{xx}$ at optimal doping ($v = 2.5$) showing a critical temperature $T_c = 5$K, which is defined by the temperature of 50% normal-state resistance. The inset shows differential resistance measurements $dV_{xx}/dI$ vs. bias current $I$, showing a sharp SC critical current transition. (**c**) Hysteresis loop reveals an orbital magnet state in device A3 at filling $v = 0.96$ and $T = 100$mK (bottom panel), which exists only in a narrow window of filling and $B_\perp$-field, which is seen by the non-zero resistance change between up and down sweeps $\Delta R_{xx}$ (top panel).

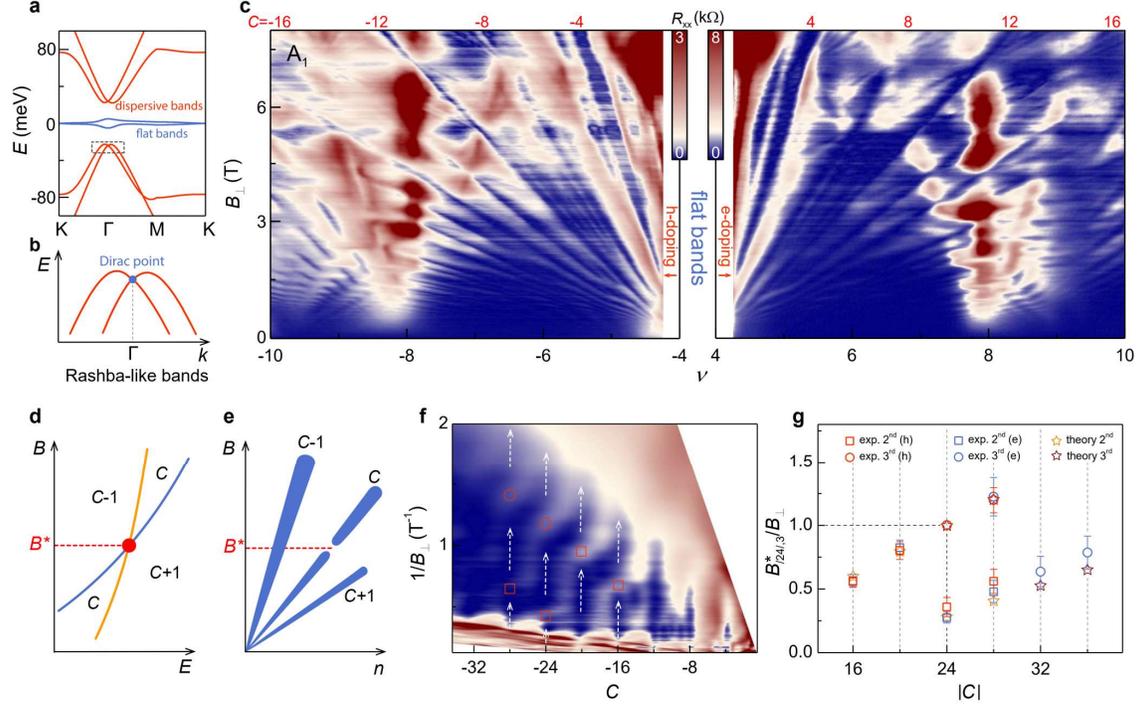

**Fig. 3. Rashba-like bands and Landau level crossings in higher energy dispersive bands.**
(**a**) Band-structure (single valley and single spin) of MATBG. (**b**) Zoom-in on the higher energy dispersive bands, which can be approximated by a Rashba-like energy dispersion. (**c**) Color plot of $R_{xx}$ vs. $\nu$ and $B_\perp$ measured at $T$ = 30mK, where we mark the Chern numbers of most dominant LL gaps. (**d**) Illustration of LL gap closings by two intersecting Landau levels at $B^*$. (**e**) Corresponding LL fan diagram in the $n$-$B_\perp$ plane shows interrupted LL gaps. (**f**) Color plot of $R_{xx}$ vs. $C$ and $1/B_\perp$ (with data from hole side of (**c**)), shows interrupting trajectories of the different LL gaps, marking the corresponding gap closings. (**g**) Comparison of the LL gap closings between experiment and theory, normalized to $B^*_{|24|,3}$, the field of the third-generation gap closing of the $C = \pm 24$ LL gaps (for electrons and holes respectively).

## Materials and Methods

Device fabrication

The MATBG samples were fabricated using a "cut-and-stack" method. The graphene was cut into two pieces using an AFM tip to prevent any unintentional strain during the tearing process. Then, a stamp of PC (poly-bis-phenol A carbonate) /PDMS (polydimethylsiloxane) mounted on a glass slide was used to pick up an hBN flake (typically 10-15 nm thick) at 100-110°C. The hBN flake was then used to carefully pick up the first half of the pre-cut graphene piece from the Si++/SiO$_2$ (285nm) substrate. The second layer of graphene was rotated by a target angle of $\theta$ = 1.1°-1.2° and picked up simultaneously by the hBN/graphene stack from the last step at 100°C. Subsequently, another hBN layer was picked up to completely cover the tBLG. Final layer of the heterostructure consists of a few-layer of graphite, which acts as a local back gate. In the end, the PC was melted at 180°C and the full stack was dropped on an O$_2$ plasma cleaned Si++/SiO$_2$ chip. Electrical connections to the tBLG were made by CHF$_3$/O$_2$ plasma etching and deposition of Cr/Au (6nm/50 nm) as metallic edge contacts.

Measurements

All of our transport measurements were carried out in a dilution refrigerator (BlueFors SD250) with a base temperature of 20mK. We have used a standard low-frequency lock-in technique (Stanford Research SR860 amplifiers) with an excitation frequency f = 13.111 Hz. To achieve a lower electron temperature in our measurements, we used very low excitation current (~ 10 nA) owing to the risk of overheating electrons and the fragility of superconductivity phases. Keithley 2400 voltage source meters was used to control the back gate voltage. The differential resistance d$V_{xx}$/dI was measured with a ~1 nA AC excitation current applied through a AC signal generated by the lock-in amplifier in combination with a 10 MOhm resistor. DC bias current was applied through a 1/100 divider and a 1 MOhm resistor before combining with the AC excitation. As-induced differential voltage was further measured at the same frequency of 19.111Hz with standard lock-in technique. We also performed electronic filtering of the measurement setup using a network of commercially available low-pass RC and LC filters.

Twist angle extraction

The phase diagrams shown in Extended Data Figure 3-5 are used to estimate the twist angle $\theta$ in the measured devices. We use the relation, $n_S = 8\theta^2/\sqrt{3}\,a^2$, where $a$ = 0.246 nm is the lattice constant of graphene and $n_S$ is the charge carrier density corresponding to a fully filled superlattice unit cell. Quantum oscillations propagating outside of the fully filled flat band was used to define $n_S$. The carrier density was calibrated with trajectory of LLs and low-field Hall effect. We find the twist angle in A$_1$ of $\theta \approx 1.04°$, A$_2$ of $\theta \approx 1.03°$, A$_3$ of $\theta \approx 1.10°$.

# Supplementary Text

## 1. Device information

We have measured three different devices with twist angles $\theta$ [($A_1$ =) 1.04°, ($A_2$ =) 1.03° and ($A_3$=) 1.1°] very close to magic angle. Extended Data Fig.1 shows the optical images of all the final devices.

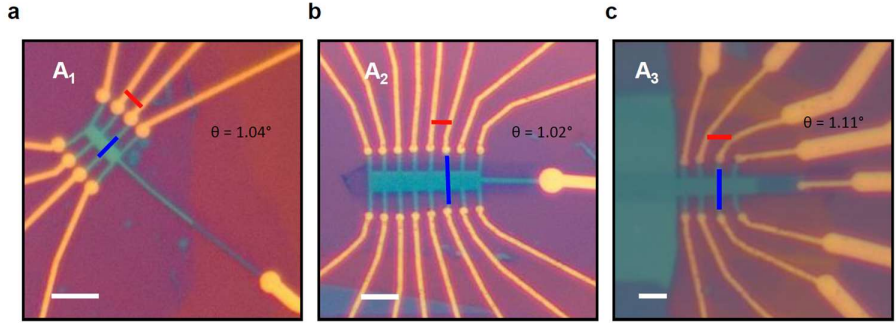

**Extended Data Fig. 1 | Optical images of the device $A_1$, $A_2$, $A_3$.** (**a**) $A_1$; (**b**) $A_2$; (**c**) $A_3$. The optical image of final devices etched in multi-terminal Hall-bar geometry from graphite/hBN/MATBG/hBN stack. The red and blue lines in the device region mark the probes used for the measurement of $R_{xx}$ and $R_{xy}$ respectively. Scale bars are 5 μm.

## 2. Hall effect:

We have also measured Hall effect at relatively lower magnetic field for all the devices. Extended Data Fig. 2 shows the Hall carrier density ($n_H$) as a function of gate voltage induced total carrier density $n$ for device $A_1$. Hall density $n_H$ was calculated from the measured transverse resistance and using the relation, $n_H = -B/eR_{xy}$. In an uniformly gated 2D electronic system, we could expect $n_H = n$. Fig. 2a exactly shows this linear region near CNP. However, as we reach the half filling of the superlattice unit cell, $n_H$ becomes almost zero hinting the formation of a gap. Beyond half-filled state $n_H$ gets reset by another slope due to the formation of a new Fermi surface. The light green and orange regions correspond to the $\nu = \pm 2, \pm 4$ states. Additionally, we have also observed the quantization of all the Chern states at higher magnetic field ($B$ = 6 T). Fig. 2b is the transverse conductivity plot as a function of carrier density. Quantization of different Chern states at different integer value further confirms their existence.

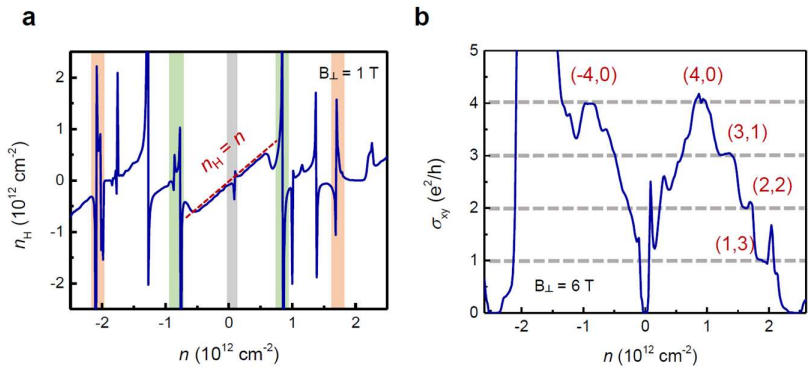

**Extended Data Fig. 2 | Hall measurement for device $A_1$.** (**a**) Hall density ($n_H$) as a function of total carrier density ($n$) extracted from Hall measurement at $B$ =1 T. (**b**) Transverse conductivity $\sigma_{xy}$ ($e^2/h$) as a function of carrier density. Plateaus correspond to the quantization of different Chern insulators.

## 3. Landau fan diagram of MATBG inside the flat band:

Extended Data Fig. 3-5 demonstrate the Chern insulating states combined with the Landau levels inside the flat band of MATBG. Similar as the spectrum in main Fig. 1b, all the devices show clear evidence of Chern insulator. Device $A_1$ has prominent Chern insulators with Chern number $C = \pm 4, 3, 2, 1$ emerging from the subsequent superlattice unit cell filling of $\nu = 0, +1, +2, +3$ without interrupted by any Landau gaps. Since we did not observe Landau levels in this device even at lowest temperature, signature of Chern states confirms their robust topological origin. However, device $A_2$ and $A_3$ has the similar Chern states along with the Landau levels emerging from CNP and $\nu = +2$ state. $A_3$ shows all the above-mentioned Chern states accompanied by $C = -2$ insulators emerging from $\nu = -2$. In addition, it has a new set of Landau levels emerging from $\nu = \pm 2$ states confirming the formation of new Fermi surface. In devices $A_2$ and $A_3$ we observed signature of several Chern insulators ($C = \pm 1, \pm 2, \pm 3$) emerging from a single filling, $\nu = \pm 3$, which hints the competition between different Chern numbers. However, they are not fully developed due to the interruption by $\nu = \pm 3$ interaction induced gap. We also observed a signature of orbital magnet at $\nu = +1$ state in device $A_3$ (will be discussed later).

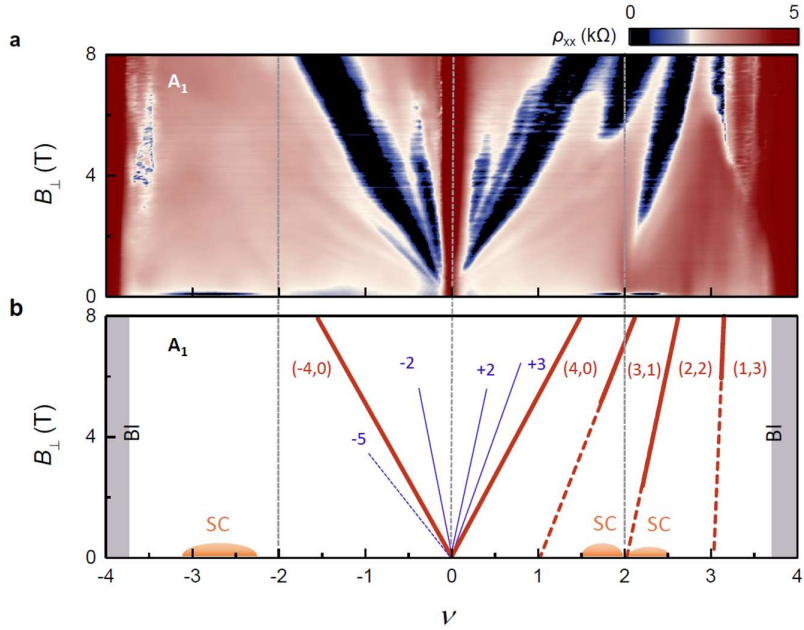

**Extended Data Fig. 3 | Full range magnetic field phase diagram of flat band in device $A_1$ at 30 mK.** (**a**) Longitudinal resistance (color plot) as a function of filling factor, $\nu$ and magnetic field $B_\perp$. (**b**) Schematics of the Landau level and Chern insulators shown in (**a**). Dark orange lines correspond to the Chern insulators $(C, \nu) = (\pm 4, 0), (3, 1), (2, 2), (1, 3)$. Blue lines are the Landau levels emerging from CNP. Light orange regions define the position of superconductors in the phase space.

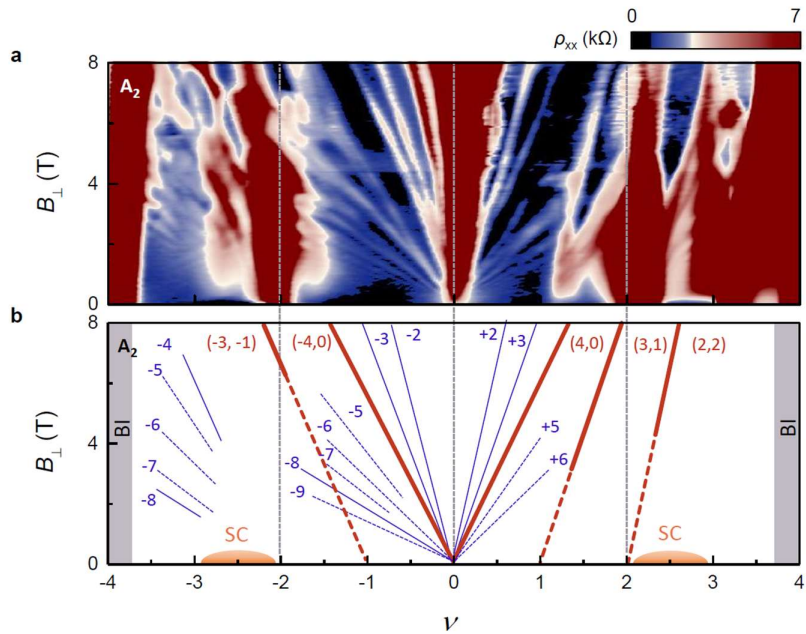

**Extended Data Fig. 4 | Full range magnetic field phase diagram of flat band in device A₂ at 30 mK.** (**a**) Longitudinal resistance (color plot) as a function of filling factor, ν and magnetic field $B_\perp$. (**b**) Schematic image of the Landau level and Chern insulators shown in (**a**). Dark orange lines correspond to the Chern insulators $(C, \nu) = (\pm 4, 0), (\pm 3, \pm 1), (2, 2)$. The blue lines correspond to the different Landau levels emerging from ν = 0, -2. Light orange regions mark the superconductors.

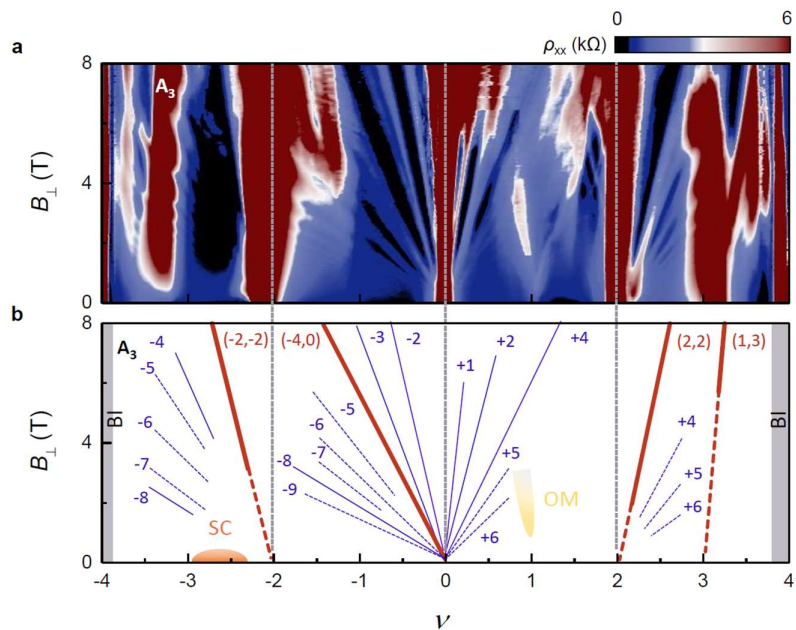

**Extended Data Fig. 5 | Full range magnetic field phase diagram of flat band in device A₃ at 30 mK.** (**a**) Longitudinal resistance (color plot) as a function of filling factor, ν and magnetic field $B_\perp$. (**b**) Schematic of all the phases. Dark orange lines correspond to the Chern insulators $(C, \nu) = (-4, 0), (\pm 2, \pm 2), (1, 3)$. The blue lines correspond to the different Landau levels emerging from ν = 0, ±2, -3. Light yellow region mark the position of orbital magnet originating from ν = 1.

### 4. Gap extraction for Chern insulator:

Extended Data Fig. 6 shows the temperature dependence of both longitudinal and transverse resistance $R_{xx}$, $R_{xy}$ of the Chern insulators. We have measured the Chern states at several temperature starting from 20 mK till 10 K at different magnetic field. Upper panel of Fig. 6 a, b, c, d are the plot of $R_{xx}$ as a function of filling factor ν at 6T for the Chern states (4,0), (3,1), (2,2), (1,3). We have observed that the $R_{xx}$ goes to zero because of the quantization of the state. Lower panel is the transverse conductivity $\sigma_{xy}$ as a function of filling factor ν. They are perfectly quantized at $4e^2/h$, $3e^2/h$, $2e^2/h$, $e^2/h$ respectively. Finally, we calculated the gap of all the Chern states from the temperature activation behavior, $R_{xx} \sim e^{-\Delta/2T}$ at 6T. Fig. 6e is the longitudinal resistance $R_{xx}$ in logarithmic scale as a function of inverse temperature ($T^{-1}$). By the Arrhenius fitting (shown in the corresponding dotted straight lines) we have calculated the gaps for each Chern state. Strikingly, these gaps $\Delta_{(4, 0)} = 11.2$K, $\Delta_{(3, 1)} = 7.1$ K, $\Delta_{(2, 2)} = 10.4$ K, $\Delta_{(1, 3)} = 4.4$K are much higher than the typical Landau gaps in MATBG. This allows us to disentangle Chern states from typical Landau levels in the system.

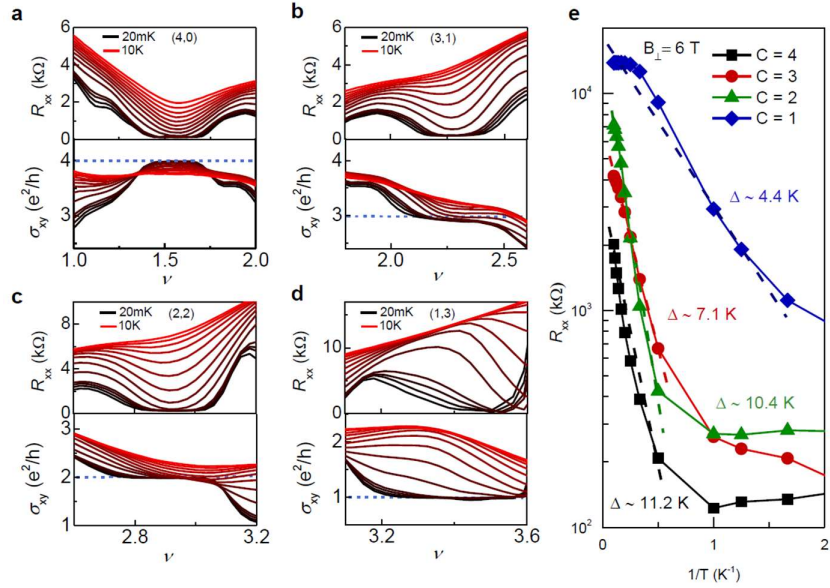

**Extended Data Fig. 6 | Temperature dependence and gap extraction for all the Chern insulators.** (**a**), (**b**), (**c**), (**d**) upper and lower panels shows the longitudinal resistance, $R_{xx}$ and transverse conductivity $\sigma_{xy}$ of the Chern insulator (4,0), (3,1), (2,2) and (1,3) as a function of ν taken at several temperatures. Lower panel corresponds to the quantization of $\sigma_{xy}$ below 600mK. (**e**) Extraction of gap for all the Chern insulators. Longitudinal resistance $R_{xx}$ is plotted as a function of inverse temperature ($T^{-1}$) at 6 T. The straight dashed lines correspond to the fitting for $R_{xx} = e^{-\Delta/2T}$. The data is measured for device **A₁**.

### 5. Superconductivity and correlated insulators:

In all of our devices, we observed several superconducting domes at different fillings along with correlated insulating states sometimes. Main Fig. 1e collectively represents all the SC domes in different devices. Extended Data Fig. 7a shows longitudinal resistance $R_{xx}$ as a function of filling factor for several temperatures from 100 mK to 40 K. Correlated insulating states appeared at ν = ±2 states accompanied by two strong superconducting domes. Fig. 7b corresponds to the temperature dependence of the resistance at optimal doping of the superconduc-tor. In this device $A_2$ we have observed a critical temperature of 5 K which is the highest recorded $T_c$ in MATBG devices so far. We further measured the differential conductance $dV_{xx}/dI$ as a function of perpendicular magnetic field and d.c. current bias, $I_{dc}$. Fraunhofer like pattern confirms the existence of superconductivity in our devices.

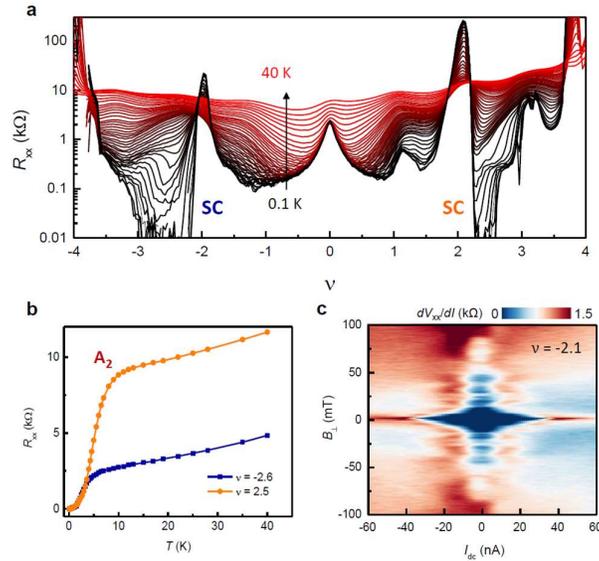

**Extended Data Fig. 7 | Superconductivity in device $A_2$.** (**a**) Longitudinal resistance $R_{xx}$ vs. $\nu$. (**b**) Temperature dependence of longitudinal resistance $R_{xx}$ at optimal doping of the superconducting dome. (**c**) Differential resistance $dV_{xx}/dI$ as a function of DC current $I_{dc}$ and $B_\perp$ showing Fraunhofer patterns.

### 6. Magnetism:

In addition to Chern insulators and superconductors we also have observed a magnetic ordered state in device $A_3$ close to $\nu = +1$ state. At zero magnetic field we did not observe any peak in $R_{xx}$. Above a magnetic field of 1 T a strong hysteretic increase of $R_{xx}$ appear. As we measured $R_{xx}$ as a function of upward and downward magnetic field, a clear hysteretic behavior was observed indicating the formation of a magnetic state. The hysteretic behavior faded out above $T = 750$ mK. We further measured the differential resistance as a function of $B_\perp$ and d.c. excitation current, which shows a critical current for this state pointing to a typical phase transition.

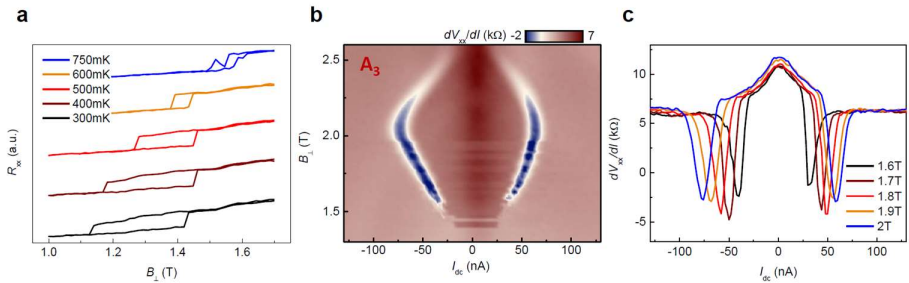

**Extended Data Fig. 8 | Orbital magnetism close to (3,1) state in $A_3$.** (**a**) Temperature dependence of the hysteresis loop of longitudinal resistance $R_{xx}$ as a function of $B_\perp$. (**b**) Color map of differential resistance $dV_{xx}/dI$ vs. DC current $I_{dc}$ and $B_\perp$ with carrier density fixed at $\nu = +1$, showing the critical current of the magnetic state. (**c**) Line-cut of $dV_{xx}/dI$ in (**b**) at different $B_\perp$, with critical current indicated by the deeps of $dV_{xx}/dI$.

### 7. Hofstadter spectrum of MATBG outside the flat band:

We further tuned the carrier density in MATBG beyond the flat band to study the higher order dispersive bands. Very low charge and twist angle disorder allows us to clearly resolve the Landau levels emerging from Rashba-like dispersive band at low magnetic field. Extended Data Fig. 9a represents a full range magnetic field phase space of device $A_1$. Apart from all the prominent Chern insulators inside the flat band, we observed two sets of Landau levels emerging from $\nu = \pm 4$ which originated from the dispersive band. The Landau gaps from one band get interrupted by another set of Landau levels from the other band. Different Landau level crossings have been discussed thoroughly in the main text. Moreover, the theoretical calculation of Rashba-like dispersive band has been addressed in the next part of Supplementary text. We found that experimental value of different gap closings match exactly with the theoretical prediction by a parameter independent factor. For the first time, this allows us to define the band structure of MATBG at higher energy. To confirm the Chern number of a state we also have measured the transverse conductance $\sigma_{xy}$ at 6T. From the quantized plateau, we assigned a Chern number to them. Fig. 9b shows $\sigma_{xy}$ and $\rho_{xx}$ as a function of $\nu$. Light orange, green and yellow regions mark the quantized plateau at $4e^2/h$, $8e^2/h$, and $12e^2/h$ respectively, which correspond to the Landau level marked in same color in Fig. 9a.

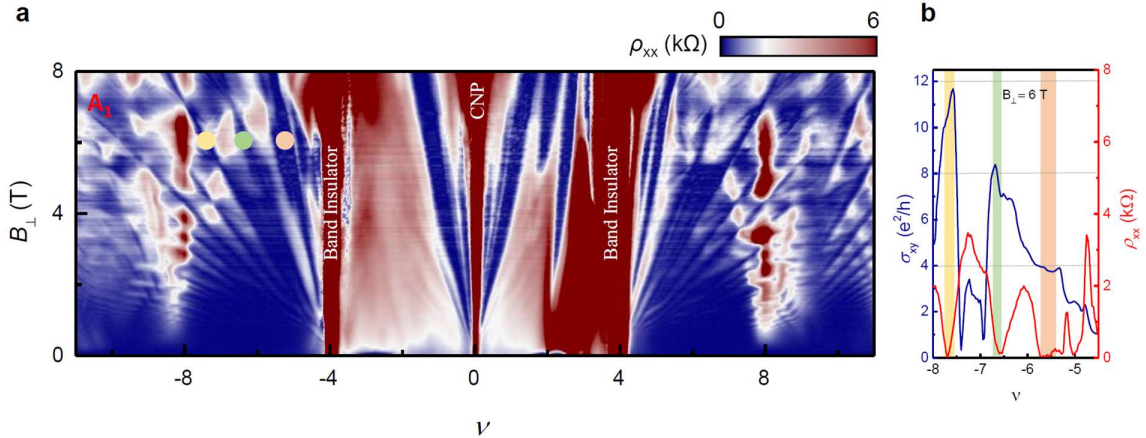

**Extended Data Fig. 9 | Full range magnetic field phase diagram of the higher energy dispersive band in device $A_1$ at 30 mK.** (**a**) Longitudinal resistance (color plot) as a function of filling factor, $\nu$ and magnetic field, $B_\perp$. (**b**) $\sigma_{xy}$ and $\rho_{xx}$ as a function of $\nu$ shows clear quantization of Chern states outside the flat band.

### 8. Theoretical Derivation of Rashba Landau Levels:

We model the low energy behavior of the two passive bands in the Bistritzer-MacDonald Hamiltonian (see Fig. 3A Of the Main Text) with a two band Rashba-like model, defined by

$$H = \frac{k_x^2 + k_y^2}{2m}\sigma_0 + \lambda(k_x\sigma_1 + k_y\sigma_2)$$

which describes the low energy theory of a Dirac node (with velocity λ) which interpolates to the free electron dispersion with effective mass m. Here we have set $\hbar = 1$. In App. 9, we discuss a perturbative derivation of the Rashba form from the Bistritzer-MacDonald Hamiltonian. By canonically coupling this model to a magnetic field, $k_x \to p_x \equiv -i\partial_x + A_x$ and $k_x \to p_y \equiv -i\partial_y + A_y$ (setting e = 1), we find

$$[p_x, p_y] = -i(\partial_x A_y - \partial_y A_x) = -iB \ .$$

We follow the well-known procedure of defining the raising and lowering operators $a = (p_x + ip_y)/(2B)^{\frac{1}{2}}$ and $a^\dagger = (p_x - ip_y)/(2B)^{\frac{1}{2}}$ which satisfy $[a,a^\dagger] = 1$ and $a^\dagger a = (k_x^2 + k_y^2 - B)/(2B)$. The vacuum $|0\rangle$ is defined such that $a|0\rangle = 0$. Hence we can rewrite the Hamiltonian as

$$H = \begin{pmatrix} \frac{B}{2m}(2a^\dagger a + 1) & \lambda\sqrt{2B}a \\ \lambda\sqrt{2B}a^\dagger & \frac{B}{2m}(2a^\dagger a + 1) \end{pmatrix} \ .$$

The Landau level spectrum is derived by acting H on the Fock states:

$$H \begin{pmatrix} |\ell - 1\rangle \\ |\ell\rangle \end{pmatrix} = \begin{pmatrix} \frac{B}{2m}(2(\ell-1) + 1) & \lambda\sqrt{2B\ell} \\ \lambda\sqrt{2B\ell} & \frac{B}{2m}(2\ell + 1) \end{pmatrix} \begin{pmatrix} |\ell - 1\rangle \\ |\ell\rangle \end{pmatrix}, \qquad \ell \geq 1$$

where we have used that $a|\ell+1\rangle = (\ell+1)^{1/2}|\ell\rangle$ and $a^\dagger|\ell\rangle = (\ell+1)^{1/2}|\ell+1\rangle$. There is also a zero-mode given by

$$H \begin{pmatrix} 0 \\ |0\rangle \end{pmatrix} = \begin{pmatrix} 0 & 0 \\ 0 & \frac{B}{2m} \end{pmatrix} \begin{pmatrix} 0 \\ |0\rangle \end{pmatrix} \ .$$

By diagonalizing the representations of the Hamiltonian on the Fock states, we find the Landau levels

$$E_{+,\ell}(B) = \frac{1}{m}\left(\ell B + \sqrt{\frac{B^2}{4} + \xi B\ell}\right), \quad \ell = 0, 1, \ldots$$

$$E_{-,\ell}(B) = \frac{1}{m}\left(\ell B - \sqrt{\frac{B^2}{4} + \xi B\ell}\right), \quad \ell = 1, 2, \ldots$$

where $\xi = 2m^2\lambda^2$ is the coupling parameter, and $1/m$ is an overall scale.

We now discuss the degeneracy of each Landau level. In a finite sample of area $A = N\Omega$ where N is the number of moiré unit cells, each with area $\Omega$, semi-classical quantization gives a total number of states $N = BA/2\pi = \varphi N/2\pi$ where $\varphi = B\Omega$ is the flux through a single Moire unit cell. We will find it useful to choose units where $\Omega = 1$ so that $B = \varphi$. Then the density of states per unit cell is $n = \varphi/2\pi$. If C Landau levels are filled, then each of the C levels contributes to the density of states and we recover the Streda formula $n = C \varphi/2\pi$, consistent with the fact that each Landau band carries a Chern number $C = 1$. When the resistance is plotted as a function of $\varphi$ and n, the Streda formula causes gaps of Chern number C to appear as distinctive linear features, the so-called Landau fan. In the case of a free particle with $\lambda = 0$, gaps of all Chern numbers appear at every flux and never close (for small field below the Hofstaeder regime), so the lines in the Landau fan are never broken. We will now see that taking $\lambda$ nonzero causes gap closings at critical values of the magnetic field. At these points, we expect to see low resistance peaks interrupting the lines of the Landau fan observed in the experiment, as shown in Fig. 10.

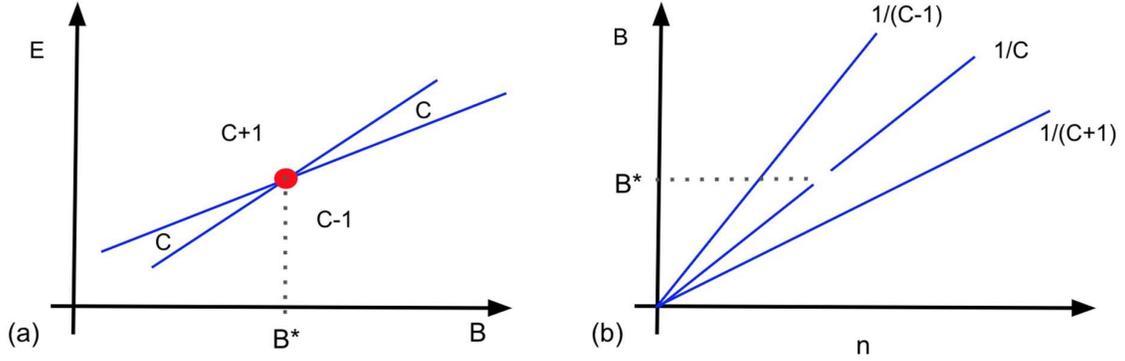

**Extended Data Fig. 10 | Schematics of LL crossings.** We illustrate how a level crossing in the Landau level spectrum affects the Landau fan. (**a**) We show two Landau levels each carrying Chern number $C = 1$ intersecting at a critical magnetic field $B^*$. For generic $B$, there will be gaps with Chern number *C-1*, *C*, and *C+1*. At $B^*$ exactly, there are only *C+1* and *C-1* gaps because the *C* gap closes. (**b**) Here we show the Landau fan, which show how the gaps of Chern number *C* change their carrier density *n* as the magnetic field is varied. The Streda formula guarantees linear trajectories with slope *1/C* for a gap of Chern number *C*. However, at $B^*$ where the gap closes, we expect a break in the trajectory where the bulk conductivity is high. Under experimental conditions, this gap will be broadened into a finite region.

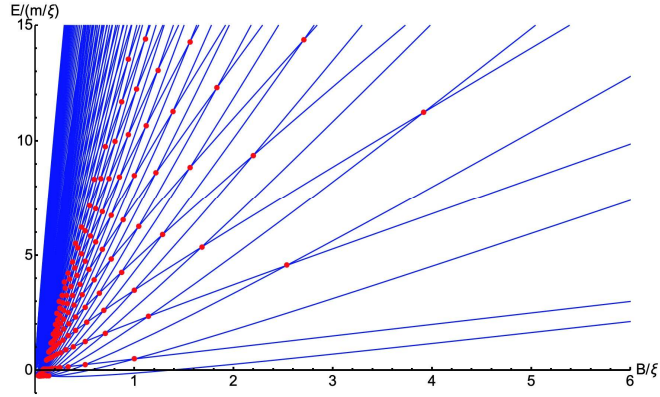

**Extended Data Fig. 11 | Calculated Landau level spectrum.** We show the Landau levels of the Rashba Hamiltonian, and indicate the level crossings with red dots. These level crossings appear in "magic series" that we obtain expressions for at the end of this section.

In order to discuss the band crossings, we will make another rescaling and let B = ξb where b is a dimensionless field strength. Then we can rewrite the Landau levels as arising from two branches

$$E_{\pm,\ell}(b) = \frac{\xi}{m}\left(\ell b \pm \sqrt{\frac{b^2}{4} + b\ell}\right), \quad \ell = \begin{cases} 0, 1, \ldots, & + \\ 1, \ldots, & - \end{cases}$$

with all the parameter dependence factored into an overall energy scale. As such, the critical b values where the Landau levels intersect are independent of all the parameters of the Hamiltonian. The Landau levels and their level crossings are shown in in Fig. 11. One may verify algebraically that only three types of crossings can occur: intersections of the $E_{+,\ell}(b)$ and $E_{-,\ell'}(b)$ levels for $\ell' \geq \ell + 2$ (denoted by $b_+(\ell, \ell')$), intersections of $E_{-,\ell}(b)$ and $E_{-,\ell'}(b)$ (denoted by $b_-(\ell, \ell')$), and intersections of $E_{-,\ell}(b)$ $E_0(b)$. These crossings occur at

$$b_\pm(\ell, \ell') = \frac{1}{\ell + \ell' \mp \sqrt{4\ell\ell' + 1}} \quad \text{with} \quad \begin{cases} \ell' \geq \ell + 2, & + \\ \ell' \neq \ell, & - \end{cases}$$

which express the crossings in terms of the Landau level index. Using this expression, we will now determine the succession of level crossings which interrupt gaps of Chern number C, as observed experimentally in Fig. 3C of the Main Text. We being by studying the large b limit, where the Landau levels take the form

$$E_{\pm,\ell}(b) = \frac{\xi}{m}\left(b(l \pm 1/2) \pm l + O(1/b)\right).$$

We see directly that $E_{+,\ell}(b) - E_{-,\ell+1}(b) = 2\ell + 1 + O(1/b)$ and $\partial_b E_{+,\ell}(b) = \partial_b E_{-,\ell+1}(b)$ to $O(1/b^2)$. We also see that the ordering of levels is $E_{-,1} < E_{+,0} < E_{-,2} < E_{+,1} < E_{-,3} < E_{+,2} < \ldots$ . The levels are organized into pairs of "parallel levels" $E_{-,\ell+1}$ and $E_{+,\ell}$ which are gapped from each other and have the same slope to leading order in b. We can organize the large b spectrum into two kinds of gaps, those between the parallel levels $E_{+,\ell}$ and $E_{-,\ell+1}$ (hence having an odd number of Landau levels below), and those separating neighboring pairs of parallel levels (hence having an even number of Landau levels below). Recall that $b = B/\xi$ is dimensionless, so the physical meaning of large b depends on the Rashba coupling. As we will prove shortly, the critical value of b for which the largest-b gap of Chern number C closes depends on C. For C=4, Fig. 11 gives b ~2.5 (we will shortly give an exact expression.) From experiment, we find $\xi \sim .18$ corresponding to a value of $\varphi \sim .45$, or B ~ 10 T. This is larger than can be currently reached in experiment, but smaller fields gaps closings, to be discussed after, are observable.

We now want to determine where these gaps first close as b is decreased from infinity. We first consider the even Chern number gaps of $C = 2\ell + 2$, $l = 0, 1, \ldots$ between $E_{+,\ell}(b)$ and $E_{-,\ell+2}(b)$. The two levels $E_{+,\ell}(b)$ and $E_{-,\ell+2}(b)$ are not parallel and the gap closes when they intersect at $b_+(\ell, \ell + 2)$. Rewriting this crossing in terms of $\ell= (C-2)/2$, we find that the initial gap closing (indexed from zero) for C = 2,4,6... is given by

$$b_{C,0} = \frac{1}{C - \sqrt{C^2 - 3}} \quad \text{for } C \geq 2, \text{ even}.$$

We now consider the odd Chern number gaps of $C = 2\ell + 1$, $\ell = 1, 2, \ldots$ in between $E_{-,\ell+1}$ and $E_{+,\ell}$. Because these levels are parallel, the gap closing happens in a manner different to the even C case. The parallel $E_{-,\ell+1}$ and $E_{+,\ell}$ levels will both be crossed by the neighboring levels of different slopes, respectively $E_{+,\ell-1}$ and $E_{-,\ell+2}$, but these crossings do not close the C $= 2\ell + 1$ Chern gap. Only once $E_{+,\ell-1}$ and $E_{-,\ell+2}$ cross at $b_+(\ell-1, \ell+2)$ does the $C = 2\ell + 1$ gap close. Writing $b_+(\ell-1, \ell+2)$ in terms of the Chern number yields

$$b_{C,0} = \frac{1}{C - \sqrt{C^2 - 8}} \quad \text{for } C \geq 3, \text{ odd}.$$

The case of C = 1 may be analyzed separately, but we will not need to do so because the C=1 gap closings appear very close to the band gap in Fig. 3 of the Main Text where the very large resistance obscures the theoretical gap closings.

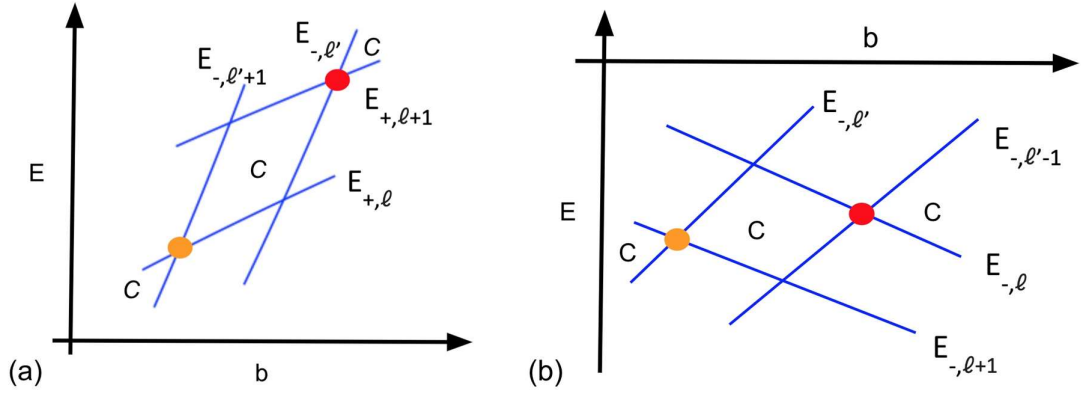

**Extended Data Fig. 12 | Structure of gap closings.** (**a**) The local structure of the next (smaller b) Chern number $C$ gap closing, shown in orange, after the level crossing of $E_{-,\ell}$ and $E_{+,\ell+1}$, shown in red. (**b**) The local structure of the series of Chern number $C$ gap closings that occur below $E = 0$. Note that the crossing are all between levels in the negative branch. These gap closings only appear for $b \leq b_-(1,2) = 1/6$.

With the large b gap closings understood, we would like extend these formulae to the cascade of additional gap closings that occur at smaller b. We will first consider the successive gaps that close due to a crossing with a level from the positive branch. In this case, the local levels for $b < b_+(\ell +1, \ell')$ appear as in Fig. 11a. The upper and lower levels are from the positive branch: $E_{+,\ell+1}$ and $E_{+,\ell}$ respectively, which must be so because the positive branch is monotone increasing in $\ell$. Note that we have assumed that $\ell \geq 0$ so $E_{+,\ell}$ is an allowed level. The right and left levels $E_{-,\ell'+1}$ and $E_{-,\ell'}$ are from the negative branch and their ordering is determined by noting that $E_{-,\ell}(b)$ is monotone increasing in $\ell$ when $E_{-,\ell}(b) > 0$. Because we assume $\ell \geq 0$ and $E_{+,\ell}(b) \geq 0$ for all nonzero b, then crossings of $E_{-,\ell'+1}(b)$ with $E_{+,\ell}(b)$ occur for $E_{-,\ell'+1}(b) > 0$. Hence we see that, if there is a gap closing for Chern number C at $b_+(\ell +1, \ell')$, the next gap closing will occur at $b_+(\ell, \ell'+1)$. By repeating this argument starting with the m = 0 gaps deduced above, we find that the *m*th gap for C = 2$\ell$ +2 occurs at $b_+(\ell-m, \ell+2+m)$ for m $\leq$ $\ell$ and the *m*th gap closing for C = 2$\ell$ +1 occurs at $b_+(\ell-1-m, \ell +2+m)$ for m $\leq$ $\ell-1$. Writing the crossings in terms of the Chern number, we find

$$b_{C, m \leq \lfloor (C-2)/2 \rfloor} = \begin{cases} \frac{C+\sqrt{C^2-(2m+1)(2m+3)}}{(2m+1)(2m+3)}, & C = 2\ell + 2, \; m = 0, 1, \ldots, \ell, \\ \frac{C+\sqrt{C^2-(2m+2)(2m+4)}}{(2m+2)(2m+4)}, & C = 2\ell + 1, \; m = 0, 1, \ldots, \ell - 1 \end{cases}$$

and it is direct to verify that the m = 0 case matches with the formulae for $b_{C,0}$ derived earlier. After crossing the $E_{+,0}$ level, the all further gap closings happen when different levels from the negative branch intersect. Consider the Chern number C gap closing of $E_{+,0}$ and $E_{-,\ell}$. Decreasing b, the next band to intersect $E_{+,0}$ is $E_{-,\ell+1}$, and the next band to intersect $E_{-,\ell}$ is $E_{-,1}$. The Chern number C gap closes at $b_-(1,\ell + 1)$ when $E_{-,\ell+1}$ and $E_{-,1}$, both in the negative branch, cross. At this point, all crossings appear in the structure of Fig 11b which gives the ordering of the bands. To derive this, consider a crossing between $E_{-,\ell'-1}$ and $E_{-,\ell}$. Using the explicit form of $b_-(\ell, \ell'-1)$, we find that, for decreasing b, $E_{-,\ell'-1}$ is next crossed by $E_{-,\ell+1}$ and $E_{-,\ell}$ is next crossed by $E_{-,\ell'}$. Hence, the Chern gap is closed when $E_{-,\ell+1}$ and $E_{-,\ell'}$ cross at $b_-(\ell+1, \ell')$. By repeating this argument, we find that the $n$th crossing after $E_{+,0}$ is given by $b_-(\ell + n, n)$. In terms of the total number of crossings m, we find gap closings at $b_-(m-\ell, \ell+2+m)$, $m > \ell$ for even C and at $b_-(m-(\ell-1), \ell+2+m)$, $m > \ell-1$ for odd $C > 1$. Writing the crossings in terms of the Chern number, we find

$$b_{C, m > \lfloor (C-2)/2 \rfloor} = \begin{cases} \frac{2m+2+\sqrt{4(m+1)^2+1-C^2}}{C^2-1}, & C = 2\ell+2, \ m = \ell+1, \ldots \\ \frac{2m+3+\sqrt{4(m+3/2)^2+1-C^2}}{C^2-1}, & C = 2\ell+1, \ m = \ell, \ldots \end{cases}$$

which, along with $b_{C, m \leq \lfloor(C-2)/2\rfloor}$, we refer to as a "magic series." In fact, it can easily be checked that $b_{C, m \leq \lfloor(C-2)/2\rfloor}$ and $b_{C, m > \lfloor(C-2)/2\rfloor}$ are dual to each other under $\lfloor(C-2)/2\rfloor \leftrightarrow m$, although we will not make use of this property. The gap closings calculated with the magic series are overlaid with the Landau level spectrum in Fig. 3C of the Main Text.

The magic series allow us to determine all the gaps $m = 0, 1, \ldots$ in the Chern number C Landau fan. As noted earlier, all the parameters of the Rashba Hamiltonian factored out into an overall energy scale and magnetic field scale. Hence, all ratios of the terms in the magic series are parameter-independent, allowing for extremely stringent tests of the Rashba approximation. Remarkably, we find very good agreement with the experimental data, as shown in Fig. 3D of the Main Text. To find $\xi$, the scale of the magnetic field, it is convenient to note that on the self-dual line $m = \lfloor(C-2)/2\rfloor$ (which maps to itself under $\lfloor(C-2)/2\rfloor \leftrightarrow m$), we calculate the simple relation

$$b_{C, \lfloor C/2 \rfloor - 1} = C^{-1} .$$

Thus when the resistance is plotted as a function of $1/B = 1/(\xi b)$ and C as in Fig. 3D of the Main Text, the gap closings in the $b_{C, \lfloor C/2 \rfloor -1}$ trajectory appear linearly with slope $\xi^{-1}$. We fit the magic series to the Landau fan data by identifying the linear trajectory and calculating $\xi$ from the slope. This fixes all of the remaining predicted gap closings, and we find excellent agreement with the experimental results. By calculating the slope of the linear trajectory on the electron side in Fig. 3D, we find the value $\xi = 1/5.3 = .186$, emphasizing that this is in units where $\hbar = e = \Omega = 1$.

### 9. Estimation of Corrugation Parameters:

Our measurement of $\xi$ from the experimental data imposes constraints on the corrugation parameters $w_0$ and $w_1$ which determine the band structure of the Bistritzer-MacDonald Hamiltonian. In principle, we can use the Bistritzer-MacDonald band structure to numerically determine $\xi(w_0, w_1)$. Our measurement of $\xi$ places a strong constraint on the allowed values of $w_0$ and $w_1$. However, we first must discuss the approximations made in applying the Rashba-like Hamiltonian. Importantly, the Bistritzer-MacDonald Hamiltonian only has $C_3$ symmetry, whereas the Rashba Hamiltonian has full rotational symmetry. Generically, we expect additional low-order terms to correct the Rashba Landau level spectrum by breaking the full rotational symmetry to three-fold symmetry. However, the excellent fit of data to the magic series leads us to believe that such symmetry-breaking terms are small, at low chemical potential in the higher-energy bands. This also guides our constraint of $w_0$ and $w_1$.

To extract and effective $\xi$ from the band structure of the Bistritzer-MacDonald Hamiltonian, we study the B=0 properties of the Rashba Hamiltonian. The band structure is given by

$$E(\mathbf{k}) = \frac{|\mathbf{k}|^2}{2m} \pm \lambda|\mathbf{k}|$$

from which we note that $|\mathbf{k}|=0$ and $|\mathbf{k}|=2m\lambda = \frac{1}{2}\xi^2$ have the same energy, $E(\mathbf{k}) = 0$. Hence the $E(\mathbf{k}) = 0$ equi-energy contour gives the value of $\xi$. For C3-symmetric perturbations, to the Hamiltonian, the average $|\mathbf{k}|$ along the contour gives an effective value of $\xi$, and the deviation of the contour $\Delta$ gives an estimate of the perturbation, as shown in Fig. 12a. We observe that the trigonal warping gets stronger for decreasing $w_1$, leading us to expect $w_1 > .95$. In this regime, $\xi$ smoothly varies as a function of $w_0$. In the electron side at filling n = +4, we find $\xi \sim .18$ leading us to estimate $.7 \leq w_0 \leq .8$. On the hole side at n = -4, we find $\xi \sim .135$ indicating a breaking of particle-hole symmetry as is expected. For simplicity, we have not incorporated particle-hole breaking in the Bistritzer-MacDonald Hamiltonian: our estimates of $w_0$ and $w_1$ are hence valid up to (small but non-negligible) particle-hole symmetry breaking effects.

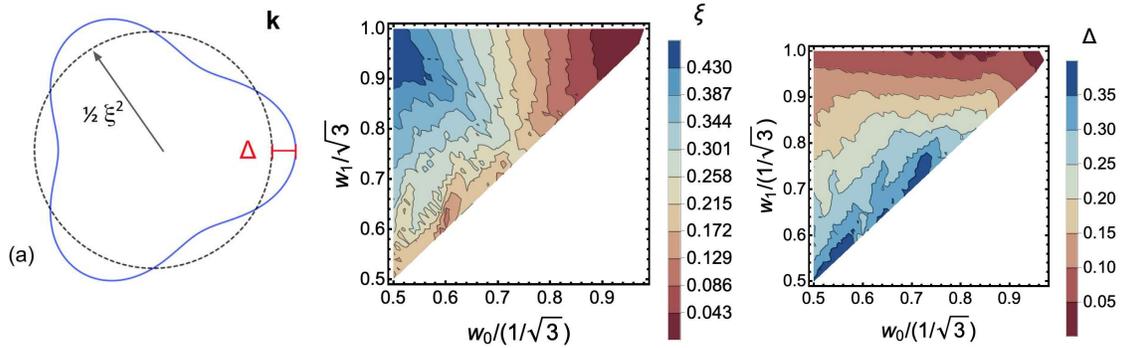

**Extended Data Fig. 12 | Band-structure parameters.** (**a**) We sketch the $E = 0$ Fermi surface for a Rashba-like Hamiltonian at zero magnetic field with a $C_3$ - symmetric perturbation. The average $|\mathbf{k}|$ gives an effective Rashba coupling $\xi$ (shown by the dashed line), and the deviation $\Delta$ (shown in red) gives an estimate of the $C_3$ - symmetric perturbation. (**b**) We plot $\xi$ as a function of $w_0$ and $w_1$. (**c**) We plot $\Delta$ as a function of $w_0$ and $w_1$. We note that $w_0 < w_1$ is imposed as a feature of the Bistritzer-MacDonald Hamiltonian that matches STM experiments (1,2).

## 10. Analytic Expressions for the Rashba Hamiltonian from Perturbation Theory

Here we study the Bistritzer-MacDonald analytically to obtain expressions for the effective Rashba Hamiltonian for the lower two bands. In Ref. 42 we invent a perturbation theory to construct the Hamiltonian of low energy bands from the MacDonald's model. The bases of MacDonald's model form an infinitely large lattice in the momentum space. In practice, one needs to set a cutoff of this lattice to obtain the Moire band structure. We find that the band dispersion of the lowest bands numerically converge to machine precision within only two shells of the lattice. Therefore, the two-shell model faithfully describe the low energy physics. We then, from this two-shell model, construct a two-band model for the lowest two bands and a two-band model for the higher two bands. Here we only give the expression of the higher two-band model: We will find that trigonal distortion terms appear to correct the Rashba form studied above, and we make simple estimates of the expected corrections in the energy spectrum.

Following this method, we obtain the Hamiltonian

$$H_{eff} = \left(E_0 + d_0(\mathbf{k})\right)\sigma_0 + \sum_{i=1}^{3} d_i(\mathbf{k})\sigma_i$$

where $E_0$ is $O(1)$, $d_0(\mathbf{k})$ is $O(\mathbf{k}^2)$, $d_i(\mathbf{k})$ are $O(\mathbf{k})$, and $\sigma_i$ are the Pauli matrices. Terms at any order may in principle be obtained in this manner. Expressions for the coefficients are as follows:

$$E_0 = \frac{1}{2}\left(\sqrt{4 + w_0^2} - \sqrt{9w_0^2 + 4w_1^2}\right),$$

$$d_1 = -\frac{w_0\left(3\sqrt{w_0^2 + 4} - \sqrt{10w_0^2 + 1}\right)}{8\left(w_0^2 + 4\right)\left(10w_0^2 + 1\right)}\left[3w_0\sqrt{w_0^2 + 4}\left(k_x - \sqrt{3}k_y\right) - \left(2\sqrt{w_0^2 + 1} + \sqrt{3}\sqrt{10w_0^2 + 1}w_0\right)\left(\sqrt{3}k_x + k_y\right)\right],$$

$$d_2 = -\frac{w_0\left(3\sqrt{w_0^2 + 4} - \sqrt{10w_0^2 + 1}\right)}{8\left(w_0^2 + 4\right)\left(10w_0^2 + 1\right)}\left[\left(2\sqrt{3}\sqrt{w_0^2 + 1} - w_0\sqrt{10w_0^2 + 1}\right)\left(\sqrt{3}k_x + k_y\right) - 3\sqrt{3}w_0\sqrt{w_0^2 + 4}\left(k_x - \sqrt{3}k_y\right)\right],$$

$$d_3 = -\frac{w_0\left(3\sqrt{w_0^2 + 4} - \sqrt{10w_0^2 + 1}\right)}{8\left(w_0^2 + 4\right)\left(10w_0^2 + 1\right)}\left[2\left(\sqrt{w_0^2 + 1}\sqrt{w_0^2 + 4}\left(\sqrt{3}k_y - k_x\right) - 6w_0\left(\sqrt{3}k_x + k_y\right)\right)\right],$$

$$d_0 = \frac{3\left(35w_0^8 - 1198w_0^6 - 6969w_0^4 - 1924w_0^2 - 76\right) + \left(37w_0^6 + 626w_0^4 - 803w_0^2 - 96\right)\sqrt{\left(w_0^2 + 4\right)\left(10w_0^2 + 1\right)}}{12\left(\left(w_0^2 + 4\right)\left(10w_0^2 + 1\right)\right)^{3/2}\left(\left(w_0^2 + 1\right)\sqrt{10w_0^2 + 1} + \left(5w_0^2 - 3\right)\sqrt{w_0^2 + 4}\right)}.$$

We remark that, in contrast to the analysis of the prior section, we work in a convention where the Brillouin zone has area $3\sqrt{3}/2$ / Ω in order to simplify the expressions. We also choose units of energy such that $v_F k_D = 1$, where $v_F$ is the Fermi velocity and $k_D$ is the Moire wavevector. Although the expressions are complicated in the chosen basis, they result in a simple zero-field spectrum given by

$$E(\mathbf{k}) = E_0 + d_0(\mathbf{k}) + \lambda_{eff}|\mathbf{k}|, \quad \lambda_{eff} = \frac{w_0\left(3\sqrt{w_0^2 + 4} - \sqrt{10w_0^2 + 1}\right)}{2\sqrt{\left(w_0^2 + 4\right)\left(10w_0^2 + 1\right)}}$$

as can be obtained by direct diagonalization. We note that because we neglect terms of order $O(\mathbf{k}^2)$ in the $d_i(\mathbf{k})$ coefficients, the effective mass is not well approximated at this order in perturbation theory. In addition, all possible $C_3$-symmetric terms at $O(\mathbf{k}^2)$ are also fully rotationally symmetric, so there is no appearance of trigonal warping at this order. We have checked higher order corrections, and we indeed see terms that break the full rotational symmetry to $C_3$ at $O(\mathbf{k}^3)$ in the spectrum. We show the comparison of the perturbative and exact spectra for $w_0 = .8/\sqrt{3}$, $w_1 = 1/\sqrt{3}$ in Fig. 13 and observe that the linear Rashba term is in good agreement, as expected at the given order in perturbation theory.

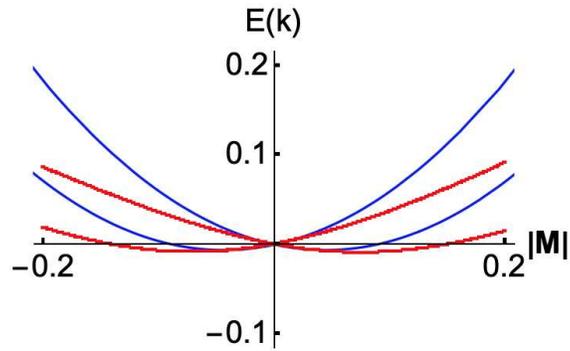

**Extended Data Fig. 13 | Rashba band-structure for different band parameters.** We plot the perturbative Rashba spectrum (blue) against the Bistritzer-MacDonald spectrum (red) along the Γ*M* line and find good agreement near the Dirac point where the Rashba coupling dominates. Because we neglect terms of O($\mathbf{k}^2$) in $d_i(\mathbf{k})$, the effective masses are not expected to match properly. We have shifted $E(\mathbf{k}=0)$ to zero for ease of comparison. We have obtained the other quadratic terms, and leave their analysis for future work.